\documentclass[%
 aip, amsmath,amssymb,
 preprint,%
]{revtex4-1}

\usepackage{graphicx}% Include figure files
\usepackage{dcolumn}% Align table columns on decimal point
\usepackage{bm}% bold math
\usepackage[utf8]{inputenc}
\usepackage[T1]{fontenc}
\usepackage{mathptmx}
\usepackage{etoolbox}
\usepackage[usenames,dvipsnames]{color}

\makeatletter
\def\@email#1#2{%
 \endgroup
 \patchcmd{\titleblock@produce}
  {\frontmatter@RRAPformat}
  {\frontmatter@RRAPformat{\produce@RRAP{*#1\href{mailto:#2}{#2}}}\frontmatter@RRAPformat}
  {}{}
}%
\makeatother
\begin{document}

%\title{Ultrafast spin dynamics in the altermagnet candidate MnTe}
\title{Time-resolved magneto-optical effects in the altermagnet candidate MnTe}

\author{Isaiah Gray}
\email{Isaiah.Gray@jhuapl.edu}
\affiliation{Department of Physics and Astronomy, University of Pennsylvania, Philadelphia, PA 19104, USA}
\address{Current address: Research and Exploratory Development Department, Johns Hopkins University Applied Physics Laboratory, Laurel, MD 20723}

\author{Qinwen Deng}
\affiliation{Department of Physics and Astronomy, University of Pennsylvania, Philadelphia, PA 19104, USA}
\author{Qi Tian}
\affiliation{Department of Physics and Astronomy, University of Pennsylvania, Philadelphia, PA 19104, USA}
\author{Michael Chilcote}
\affiliation{Materials Science and Technology Division, Oak Ridge National Laboratory, Oak Ridge, TN 37831, USA}
\author{J. Steven Dodge}
\affiliation{Department of Physics, Simon Fraser University, Burnaby, BC, Canada}
\author{Matthew Brahlek}
\affiliation{Materials Science and Technology Division, Oak Ridge National Laboratory, Oak Ridge, TN 37831, USA}
\author{Liang Wu}
\email{liangwu@sas.upenn.edu}
\affiliation{Department of Physics and Astronomy, University of Pennsylvania, Philadelphia, PA 19104, USA}

\date{\today}% It is always \today, today,

\begin{abstract}
$\alpha$-MnTe is an antiferromagnetic semiconductor with above room temperature $T_N$ = 310 K, which is promising for spintronic applications. Recently, it was reported to be an \textit{altermagnet}, containing bands with momentum-dependent spin splitting; time-resolved experimental probes of MnTe are therefore important both for understanding novel magnetic properties and potential device applications. We investigate ultrafast spin dynamics in epitaxial MnTe(001)/InP(111) thin films using pump-probe magneto-optical measurements in the Kerr configuration. At room temperature, we observe an oscillation mode at 55 GHz that does not appear at zero magnetic field. Combining field and polarization dependence, we identify this mode as a magnon, likely originating from inverse stimulated Raman scattering. Magnetic field-dependent oscillations persist up to at least 335 K, which could reflect coupling to known short-range magnetic order in MnTe above $T_N$. Additionally, we observe two optical phonons at 3.6 THz and 4.2 THz, which broaden and redshift with increasing temperature.
\end{abstract}

\maketitle

In the past several years, antiferromagnets have been developed into active elements of spintronic devices, forming memory elements \cite{WadleyScience2016, HigoNature2022}, terahertz nano-oscillators \cite{ChengPRL2016, OlejnikSciAdv2018}, and low-dissipation interconnects in hybrid quantum systems \cite{LebrunNature2018}. A natural route towards integrating these devices with traditional semiconductor functionality is to employ magnetic semiconductors \cite{JungwirthNatNano2016}; while room-temperature ferromagnetic semiconductors are rare, antiferromagnetic semiconductors are more common and offer new methods of controlling the spin degree of freedom, such as coupling between magnons and excitons \cite{BaeNature2022}. Hexagonal $\alpha$-MnTe is promising due to its high N{\'e}el temperature $T_N$ $\approx$ 310 K \cite{SzuszkiewiczPhysStatSolidi2005}, stable magnetoresistance reflecting N{\'e}el orientation \cite{KriegnerNatCommun2016}, and spin-phonon coupling \cite{ZhangJRamanSpec2020, BossiniPRB2021}. Very recently, MnTe has received increased attention after the prediction \cite{SmejkalPRX2022, SmejkalPRX2022_2} and subsequent confirmation \cite{KrempaskyNature2024} that it is not simply an antiferromagnet, but an \textit{altermagnet}. In this new magnetic phase, the spin-up and spin-down bands are neither degenerate, like conventional antiferromagnets, nor uniformly split, like ferromagnets; rather, the bands contain anisotropic spin polarization that alternates sign in momentum space while preserving zero net spin polarization.

This anisotropic spin polarization has very recently been observed via ARPES in MnTe \cite{KrempaskyNature2024}, MnTe$_2$ \cite{ZhuNature2024}, and RuO$_2$ \cite{FedchenkoSciAdv2024}, and is predicted to lead to new effects of both scientific and practical importance, such as an anomalous Hall effect \cite{SmejkalPRX2022, HayamiPRB2021}, an associated magneto-optical Kerr effect (MOKE), novel forms of spin-valley control \cite{MaNatComm2021}, and spin current generation \cite{SmejkalPRL2023, HayamiJPSJ2019}. Currently there are few experimental demonstrations of these effects, mostly in RuO$_2$ \cite{FengNatElectron2022, LiuAdvOptMater2023, BaiPRL2022, BoseNatElectron2022, FedchenkoSciAdv2024}. In MnTe, there is one report of an anomalous Hall effect attributed to altermagnetism\cite{Gonzalez_Betancourt_PRL_2023}; however, variations in stoichiometry and strain can lead to weak ferromagnetism, which can also produce an anomalous Hall effect \cite{ChilcoteAdvFuncMater2024}. Meanwhile, MOKE in MnTe has been predicted \cite{MazinPRB2023} but not yet observed\cite{ChilcoteAdvFuncMater2024}. Experimental studies of magnetism in MnTe are therefore important for better understanding of a potential altermagnetic phase.

In this letter, we perform time-resolved magneto-optical (MO) measurements in the Kerr geometry on hexagonal MnTe thin films as a function of temperature, polarization, and magnetic field. We identify optical phonon modes at 3.6 THz and 4.2 THz and show that they broaden and redshift with increasing temperature from 10 K to 350 K. In the presence of magnetic field, we observe an oscillation mode at 55 GHz below $T_N$, which we hypothesize is due to an in-plane $k = 0$ magnon,  as the oscillation is absent at zero magnetic field \cite{RezendeJApplPhys2019}. The oscillation amplitude depends on the pump polarization angle, which indicates a non-thermal origin. Surprisingly, magnetic field-dependent 55 GHz oscillations persist even above $T_N \approx$ 310 K, up to the maximum measured $T$ = 335 K, which could be due to the presence of known short-range magnetic ordering above $T_N$ in MnTe \cite{BaralMatter2022}.

\begin{figure*}
\includegraphics[width=0.8\textwidth]{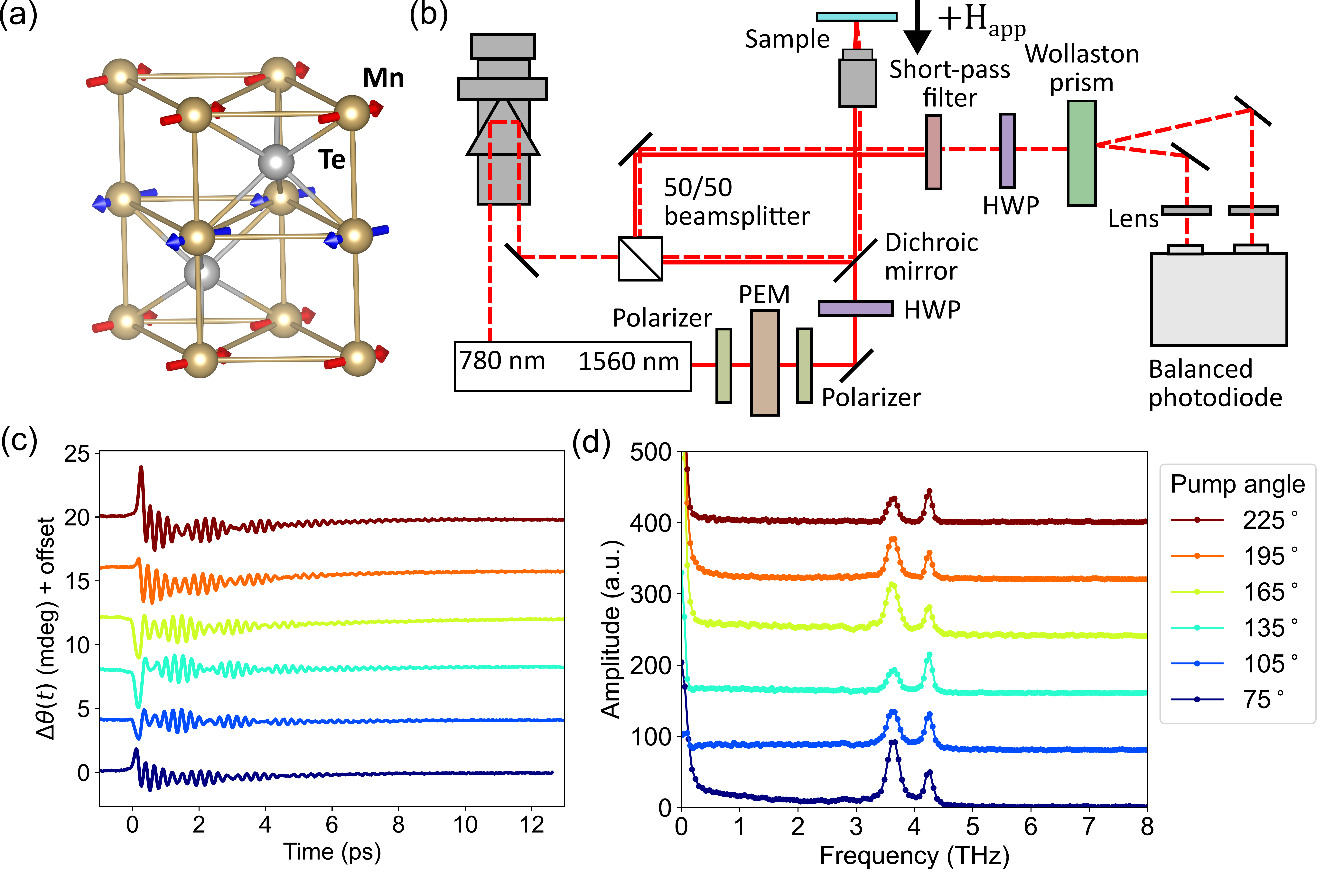}
\caption{\label{fig:Fig1} Time-resolved magneto-optical setup and observation of optical phonons in MnTe. (a) Crystal and magnetic structure of hexagonal MnTe (001). Gold and silver balls represent Mn and Te, respectively. (b) Schematic of two-color pump-probe MOKE setup with 1560 nm pump and 780 nm probe, both $\sim$100 fs in duration. (c) Probe polarization rotation $\Delta \theta(t)$ in MnTe(001) at 295 K as a function of pump polarization angle $\varphi$, and (d) associated fast Fourier transforms (FFTs). }
\end{figure*}

The structure of hexagonal $\alpha$-MnTe, with space group $P6_3/mmc$, is shown in Fig. \ref{fig:Fig1}(a). It is an A-type antiferromagnet, containing ferromagnetic planes of in-plane Mn spins that alternate spin direction along the $c$-axis. The easy axis is at 30$^\circ$ to the crystalline $a$-axis. Our samples are 40 nm-thick MnTe(001) thin films, grown on InP(111) via molecular beam epitaxy.

We perform two-color pump-probe measurements, schematically illustrated in Fig. \ref{fig:Fig1}(b). We use a 1560 nm pump at 10 mW power and a 780 nm probe at 0.7 mW power on the sample; both pump and probe pulses are $\approx$100 fs in duration and have 80 MHz repetition rate. Both beams are focused through a 10x microscope objective at normal incidence, forming a 10 $\mu$m spot size. The reflected light is short-pass filtered to reject the pump, then the polarization rotation $\Delta \theta(t)$ is measured with a Wollaston prism and a balanced photodiode. Signal-to-noise is enhanced by modulating the pump with a photoelastic modulator at 42 kHz and measuring the second harmonic of the photodiode voltage output using a lock-in amplifier. Both pump and probe are linearly polarized.

We first show short time scans (up to 13 ps) of $\Delta \theta(t)$ at room temperature (295 K) as a function of pump polarization angle $\varphi$ in Fig. \ref{fig:Fig1}(c). $\varphi$ is referenced to a InP(111) substrate crystalline axis; the MnTe crystalline axes should follow that of the substrate, likely with 60$^\circ$ crystalline twinning. After an initial sharp peak, we observe beating between two modes. Upon taking the FFTs in Fig. \ref{fig:Fig1}(d), we find peaks at 3.6 and 4.2 THz, whose amplitudes depend on $\varphi$. These frequencies match modes observed in previous Raman spectroscopy of MnTe \cite{ZhangJRamanSpec2020} that were attributed to transverse and longitudinal optical phonons, respectively. Polarization-dependent amplitudes demonstrate a non-thermal excitation mechanism, such as impulsive stimulated Raman scattering (ISRS) \cite{YanJChemPhys1985} or displacive excitation of coherent phonons (DECP) \cite{ZeigerPRB1992}. Since DECP requires absorption of a photon (meaning above-bandgap excitation), ISRS does not, and our pump wavelength is below the bandgap of MnTe(1.15 eV = 1078 nm at 295 K \cite{BossiniNewJPhys2020}), ISRS is the most likely excitation mechanism.

Our results differ from previous time-resolved reflectivity and MOKE on MnTe \cite{BossiniPRB2021, DeltenrePRB2021}, where only the doubly degenerate $E_{2g}$ phonon mode at 5.3 THz (also seen in Raman scattering \cite{ZhangJRamanSpec2020}) was observed. In these works, however, the pump wavelength was 800 nm (above the bandgap), the probe wavelength was 730 nm, and the pump-probe signal was explicitly attributed to DECP. These different excitation mechanisms and probe wavelengths might cause the differences in observed phonons.

\begin{figure*}
\includegraphics[width=\textwidth]
{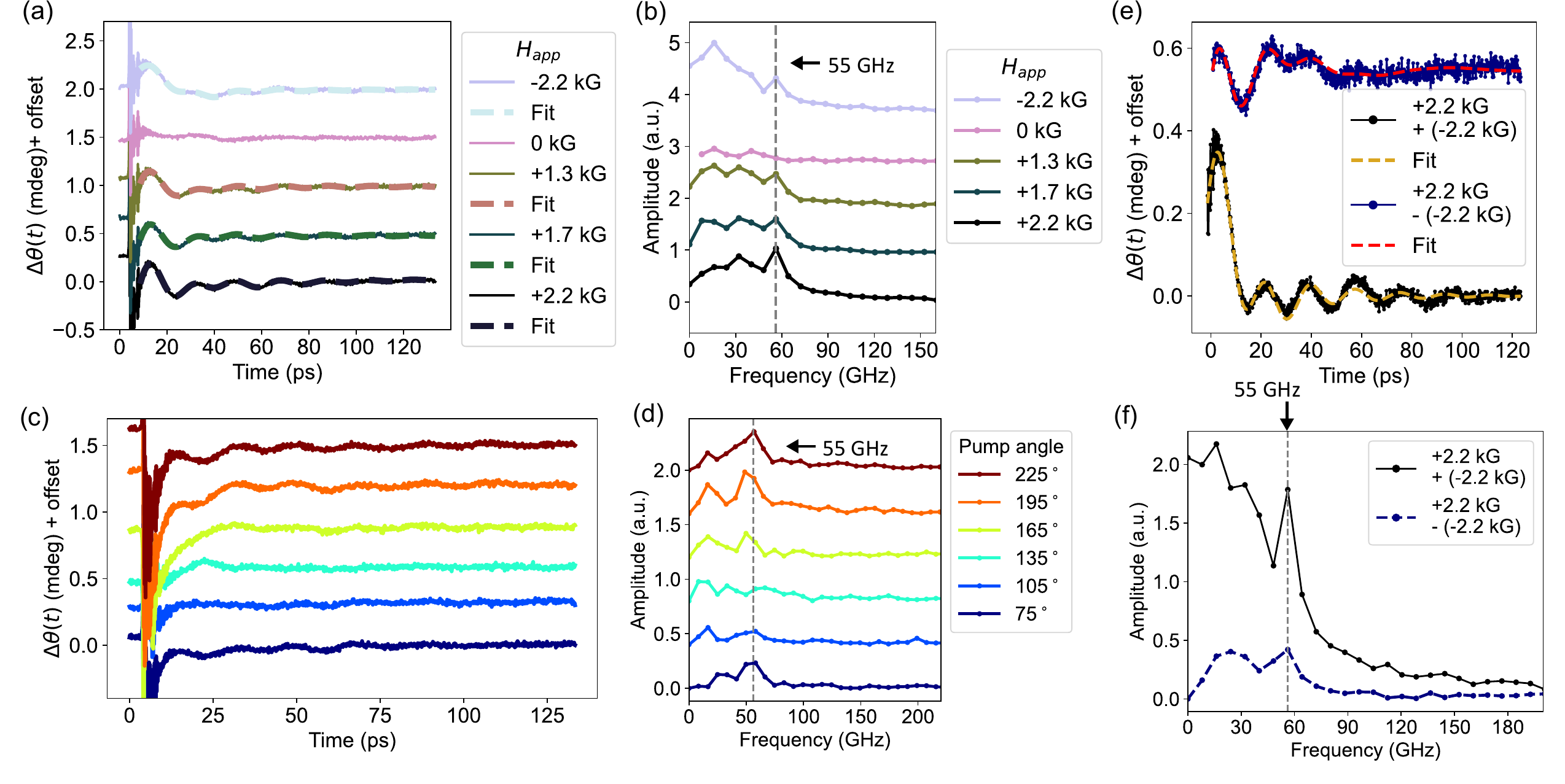}
\caption{\label{fig:Fig2} Magnon and acoustic phonon oscillations at 295 K. (a) Longer scans (133 ps duration) at fixed $\varphi = 225^\circ$ as a function of field $H_{app}$, and (b) associated FFTs. We observe a mode at 55 GHz with field-dependent amplitude, which indicates spin oscillations, as well as modes at $\sim$17 and 34 GHz that may be due to acoustic phonons. (c) Pump polarization dependence of $\Delta \theta(t)$ at fixed $H_{app} = 2.2$ kG, with FFTs in (d). Oscillation amplitudes are $\varphi$-dependent, demonstrating a non-thermal origin such as phonon-induced magnons. (e) Difference and sum of $\Delta \theta(t)$ at +2.2 kG and -2.2 kG, and (f) associated FFTs. $\Delta \theta (t)$ is largely even in $H_{app}$.}
\end{figure*}

To observe spin oscillations, we fix $\varphi = 225^{\circ}$ from horizontal, where both phonon modes are excited in roughly equal magnitude, and measure $\Delta \theta(t)$ while varying the applied magnetic field $H_{app}$. To observe MO signal from out-of-plane magnetization or potential in-plane altermagnetism \cite{Gonzalez_Betancourt_PRL_2023}, we apply out-of-plane field with a permanent magnet placed behind the sample. $\Delta \theta(t)$ scans are shown in Fig. \ref{fig:Fig2}(a) and associated FFTs are shown in Fig. \ref{fig:Fig2}(b). After the initial fast phonon oscillations, we observe slower oscillations whose amplitude depends on $H_{app}$, and do not appear when $H_{app} = 0$. Oscillations of similar frequencies were observed in time-resolved transmissivity at 800 nm pump wavelength in MnTe/SrF$_2$(111) in Ref. \onlinecite{ZhuPRMater2023}, where they were tentatively attributed to acoustic phonons \footnote{ Note that GHz-scale \textit{acoustic} phonons can be excited by laser light both above the bandgap, by the deformation potential mechanism (similar to DECP), and below the bandgap, by thermoelastic or electrostriction mechanisms (see \textit{P. Ruello and V. E. Gusev, Ultrasonics 56, 21-35 (2015)} for a review.) Thus, the observation of similar-frequency acoustic phonons in both our work and Ref. \onlinecite{ZhuPRMater2023} does not necessarily contradict the differing optical phonon measurements in our work and Ref. \onlinecite{BossiniPRB2021}}. 

We fit the data from 10 ps onwards, discarding the initial phonon modes, to the sum of an exponential decay representing electron-phonon relaxation \cite{BossiniPRB2021}, two exponentially damped sinusoids representing acoustic phonons and/or magnons \cite{GamalyProgQuantElectron2013}, and a linear background:

\begin{equation}
\begin{split}
\Delta \theta(t) = A_{exp}e^{-t/t_0} + A_1 e^{-\alpha_1 t}\sin \left(2 \pi f_1 t + \phi_1 \right)& \\
+ A_2 e^{-\alpha_2 t}\sin \left( 2 \pi f_2 t + \phi_2 \right) + B t + C
\end{split}
\end{equation}

Fits are shown with the data in Fig. \ref{fig:Fig2}(a), except at $H_{app}$ = 0 kG, where a consistent fit could not be obtained. At this pump angle (225$^\circ$), the features of the data are well-captured by two damped oscillations and fitting to three did not significantly improve the fit (see the supporting information (SI)).  Best fit parameters are shown in Table \ref{tab:table1}. We find that although both oscillations go to zero when $H_{app} = 0$, $A_1$ monotonically depends on the magnitude of $H_{app}$, which suggests a magnetic origin, whereas $A_2$ does not.

Previous studies of magnons in MnTe are scarce\cite{SzuszkiewiczPhysStatSolidi2005, MuPRMater2019}.  We can estimate the $k = 0$ magnon frequencies from the anisotropy, which in MnTe is largely easy-plane because optics probe the $k = 0$ magnons.  %Within the easy plane there are nominally three equivalent easy axes; however, MnTe on InP(111) experiences biaxial tensile strain \cite{KriegnerPRB2017}, which modifies the in-plane anisotropy in a not-well-understood manner that depends also on sample thickness.  
Since the in-plane anisotropy field $H_{Az}$ is small compared to the hard-axis (easy-plane) field $H_{Ax}$, we can use the formula for the two $k = 0$ easy-plane magnon frequencies $\omega_\alpha$ and $\omega_\beta$ \cite{RezendeJApplPhys2019}:

\begin{equation}
\begin{split}
\omega_{\alpha,\beta}^2 = \gamma^2 \{ H_E (H_{Ax}+2H_{Az}) + H_0^2 & \\
\pm \left[ 4 H_0^2 H_E (H_{Ax} + 2 H_{Az}) + H_E^2 H_{Ax}^2\right]^{1/2} \},
\end{split}
\end{equation}

where $\gamma = g \mu_B/\hbar$ is the gyromagnetic ratio and $H_E$ is the exchange field. For $H_{Ax} >> H_{Az}$, Eqn. 3 reduces to: 

\begin{equation}
\omega_{\alpha 0}^2 \approx \gamma^2 (2 H_E H_{Ax} + 3H_{app}^2), \omega_{\beta 0}^2 \approx \gamma^2 (2 H_E H_{Az} - H_{app}^2),
\end{equation}

 while for $H_{app} = 0$, regardless of the anisotropy values, the frequencies are

\begin{equation}
\omega_{\alpha 0}^2 = \gamma^2 2 H_E \left( H_{Az} + H_{A x}\right), \omega_{\beta 0}^2 = \gamma^2 2 H_E H_{Az}.
\end{equation}

Although the values of $H_{Az}$ and $H_{Ax}$ are not well-known, we can get an estimate by using the spin-flop field $H_{SF} \approx \sqrt{2 H_E H_{Az}}$, about 2 T for MnTe/InP(111) \cite{KriegnerPRB2017}. Assuming $g = 2$ for MnTe, the in-plane frequency $\omega_{\beta 0}$ evaluates to 56 GHz, consistent with our measured value. From the lack of any measurable magnetization response in MnTe with OOP field up to 6 T in Ref. \onlinecite{KriegnerNatCommun2016}, we can estimate that $H_{Ax} > 6$ T, which means that the OOP magnon frequency $\omega_{\alpha 0} >$ 176 GHz - therefore, features between 17 and 35 GHz are unlikely to be magnons. They may represent acoustic phonons, or a non-monotonic magneto-optical response to sample heating \cite{YangPRM2019}. In previous studies of the easy-plane antiferromagnet NiO, the in-plane mode was observed with a linearly polarized probe through Faraday rotation, while the out-of-plane mode was observed with a circularly polarized probe through the Cotton-Mouton effect (linear birefringence) \cite{TzchaschelPRB2017}. If similar mechanisms are active in MnTe, the higher-frequency magnon may only appear when using a circularly polarized probe, which will be the subject of future work.

Following the field dependence, we measure pump polarization dependence of the spin oscillations at fixed $H_{app}$ = 2.2 kG in Fig. \ref{fig:Fig2}(c), with associated FFTs in \ref{fig:Fig2}(d). We find that the oscillation amplitudes are pump polarization-dependent, although at the current signal-to-noise level, we cannot definitively determine the symmetry. Data at $\varphi = 75^\circ$ fits better to three oscillations, while data at $\varphi = 195^\circ$ and 225$^\circ$ seems to contain only two (see SI). This suggests that multiple acoustic phonons with different pump angle dependences may be excited by the laser.

In general, magnons measured with ultrafast optical pump-probe can be thermally or non-thermally generated. In thermal generation, laser heating temporarily changes the anisotropy field $H_A$, which reorients $H_{eff}$ and causes the spins to precess around the new $H_{eff}$ \cite{LyalinNanoLett2021}. Non-thermal magnon generation is often caused by spin-phonon coupling in magnetic semiconductors and insulators when the laser coherently excites phonons \cite{BerkNatCommun2019}, and was observed in MnTe in Refs. \onlinecite{BossiniPRB2021, DeltenrePRB2021}. In those experiments, the off-normally incident laser excited the 5.3 THz $E_{2g}$ optical phonon, causing the Te atoms to oscillate out-of-phase. This modulated the Mn-Te-Mn bond angle, in turn causing an oscillation in the weak Te-mediated superexchange $J_3$ which generated magnons at the same frequency. In our data, however, the 3.6 THz and 4.2 THz oscillations in $\Delta \theta (t)$ show no field dependence within experimental resolution (see the supplemental information). Therefore, we attribute them entirely to optical phonons. The polarization dependence in Fig. \ref{fig:Fig2}(c) is consistent with a hypothesis of acoustic phonon-induced spin oscillations at 55 GHz, driven by ISRS \cite{YanJChemPhys1985}.

\begin{table}
\caption{\label{tab:table1} Fit parameters for the data in Fig. 2(a), fit to Eqn. 1. Uncertainties on the last significant digit are given in parentheses.}
\begin{ruledtabular}
\begin{tabular}{cccccc}
 $H_{app}$ & 2.2 kG & 1.7 kG & 1.3 kG & -2.2 kG \\
\hline
$A_1$ (mdeg)& 0.108(5) & 0.063(5) & 0.039(4)  & 0.049(3) \\
$\alpha_1$ (ps$^{-1}$)& 0.023(1) & 0.023(2) & 0.016(2)  & 0.026(2)  \\
$f_1$ (GHz) & 53.9(2) & 53.3(3) &53.4(3) & 56.0(5)  \\
$\phi_1$ (deg) & 13(3) & 12(5) &5.0(1)  & -16(4)\\
$A_2$ (mdeg) &0.236(8)  & 0.22(1) & 0.32(1)  &0.345(5) \\ 
$\alpha_2$ (ps$^{-1}$) &0.084(3) & 0.078(4) &0.107(4)  &0.073(2) \\ 
$f_2$ (GHz) & 36.7(5) &36.2(7) & 32.8(6) & 18.2(3) \\
$\phi_2$ (deg) &14(2) & 16(3) &20(2) & 52(2) \\
\end{tabular}
\end{ruledtabular}

\end{table} 

In many pump-probe experiments, magnetic and non-magnetic signals - for example, from acoustic phonons or phonon-induced spin tilting \cite{LatteryThesis2020, ZhangSciAdv2020} - are separated by taking the sum and difference of data acquired at opposite magnetic fields \cite{McGillPRL2004}. Therefore, in Fig. \ref{fig:Fig2}(e) we plot the sum and difference of $\Delta \theta(t)$ at $H_{app} = +2.2$ kG and -2.2 kG, with FFTs in \ref{fig:Fig2}(f). We find greater amplitude in the sum than the difference, and in fact the difference nearly cancels out after 40 ps. This means that although $\Delta \theta(t)$ is field-dependent, it does not change sign with $H_{app}$ (which can be seen as well in Fig. \ref{fig:Fig2}(a)). $\Delta \theta(t)$ that is even in $H_{app}$ can occur if $\Delta \theta (t)$ is due to a Voigt effect \cite{TzchaschelPRB2017} or optical birefringence \cite{XuPRB2019}, which is quadratic in magnetization $m$, rather than the Kerr effect that is linear in $m$. Since the net $m$ is zero in antiferromagnets, the dominant magneto-optical effect is often these quadratic terms. Static Kerr and circular dichroism measurements of these samples showed no hysteresis \cite{ChilcoteAdvFuncMater2024}, which is consistent with a lack of observed TR-MOKE signal.

\begin{figure}
\includegraphics[width=\columnwidth]
{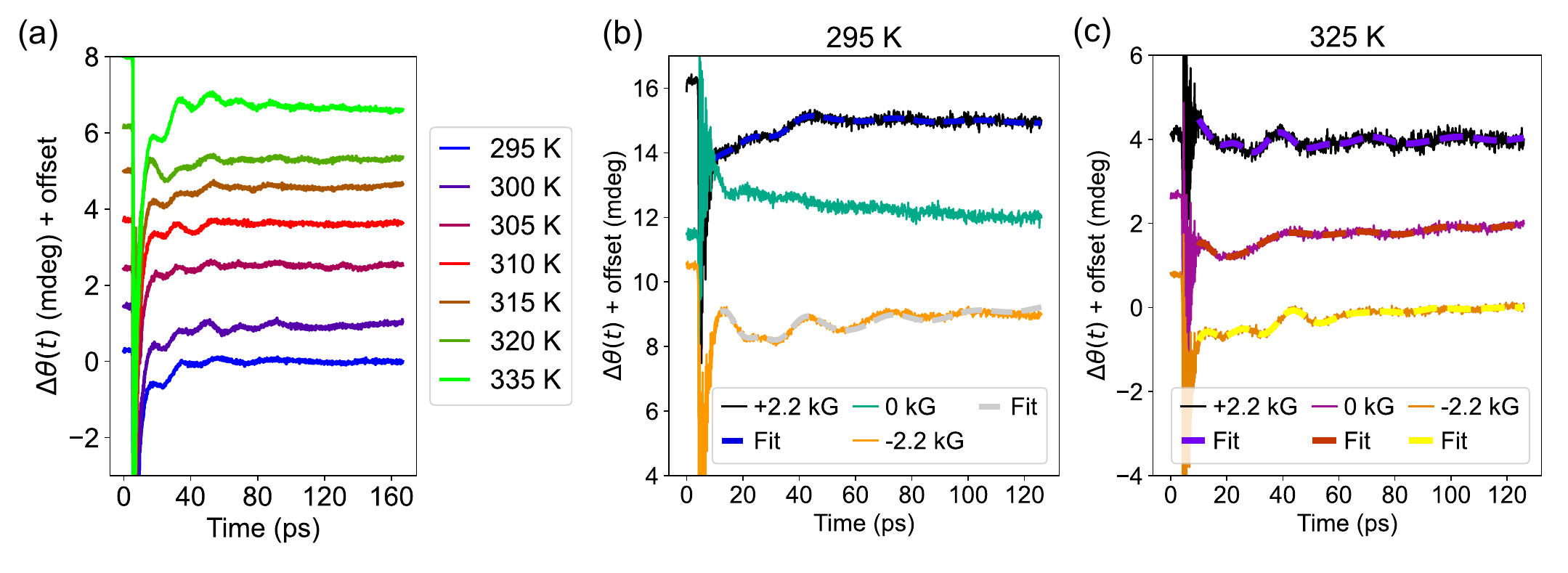}
\caption{\label{fig:Fig3} Measuring $\Delta \theta (t)$ through $T_N$. (a) $\Delta \theta(t)$ at fixed $H_{app}$ = +2.2 kG from 295 K to 335 K. 55 GHz oscillations appear at all temperatures. (b), (c) $\Delta \theta (t)$ at $H_{app}$ = +2.2 kG, 0 kG, and -2.2 kG OOP at 295 K and 325 K, respectively. At both temperatures, the 55 GHz mode appears only when $H_{app} \neq 0$, which indicates that it is magnetic. }
\end{figure}

After field and polarization dependence, we measure $\Delta \theta (t)$ as a function of temperature through $T_N$ ($\approx$ 300 K in our thin films \cite{ChilcoteAdvFuncMater2024}, slightly lowered from bulk $T_N$ = 310 K due to strain), where long-range ordering disappears.
 %Because isolating magnetic from non-magnetic contributions to $\Delta \theta (t)$ cannot be straightforwardly done by taking the difference and sum at low fields, we instead measure above $T_N$ ($\approx$ 300 K in our thin films \cite{ChilcoteAdvFuncMater2024}, slightly lowered from bulk $T_N$ = 310 K due to strain), where we expect spin oscillations to disappear.% 
 Fig. \ref{fig:Fig3}(a) shows $\Delta \theta(t)$ at fixed $H_{app}$ = +2.2 kG from 295 K to 335 K. Surprisingly, oscillations at 55 GHz seem to persist at least up to 335 K, the highest temperature measured. To determine if they are still magnetic above the nominal $T_N$, we compare $\Delta \theta (t)$ taken at $H_{app}=$ +2.2 kG, 0 kG, and -2.2 kG at 295 K and 325 K in Fig. \ref{fig:Fig3}(b) and (c), respectively. Fit parameters are shown in the SI. At both temperatures, the 55 GHz mode appears \textit{only} when $H_{app} \neq 0$, which indicates that it is still magnetic. Previous studies in both ferromagnets \cite{QinPRL2017} and antiferromagnets \cite{BiesnerPRB2022} have shown that in materials with short-range ordering above the Curie or N{\'e}el temperatures, respectively, magnetic anisotropy can persist above $T_C$ or $T_N$, and therefore collective spin oscillations can be observed at the same frequency through the transition. Since short-range magnetic order is known to exist in MnTe well above $T_N$ \cite{BaralMatter2022}, it is therefore plausible that MnTe continues to exhibit spin oscillations for some temperature range above $T_N$.
 
 %Somewhat surprisingly, oscillation components at 55 GHz seem to \textit{increase} with temperature and persist at least up to 335 K, the highest temperature measured. Separate measurements on the same sample in the SI show that the 55 GHz oscillation is still magnetic field-dependent at 325 K and disappears when $H_{app} = 0$, which would indicate that the magnon persists above $T_N$. In both ferromagnets \cite{QinPRL2017} and antiferromagnets \cite{BiesnerPRB2022}, magnons have been observed above the nominal transition temperature due to coupling to short-range order. Since short-range magnetic order is known to exist in MnTe \cite{BaralMatter2022} well above room temperature, it is reasonable that magnons would also persist for some temperature range above $T_N$.}

\begin{figure*}
\includegraphics[width=\textwidth]
{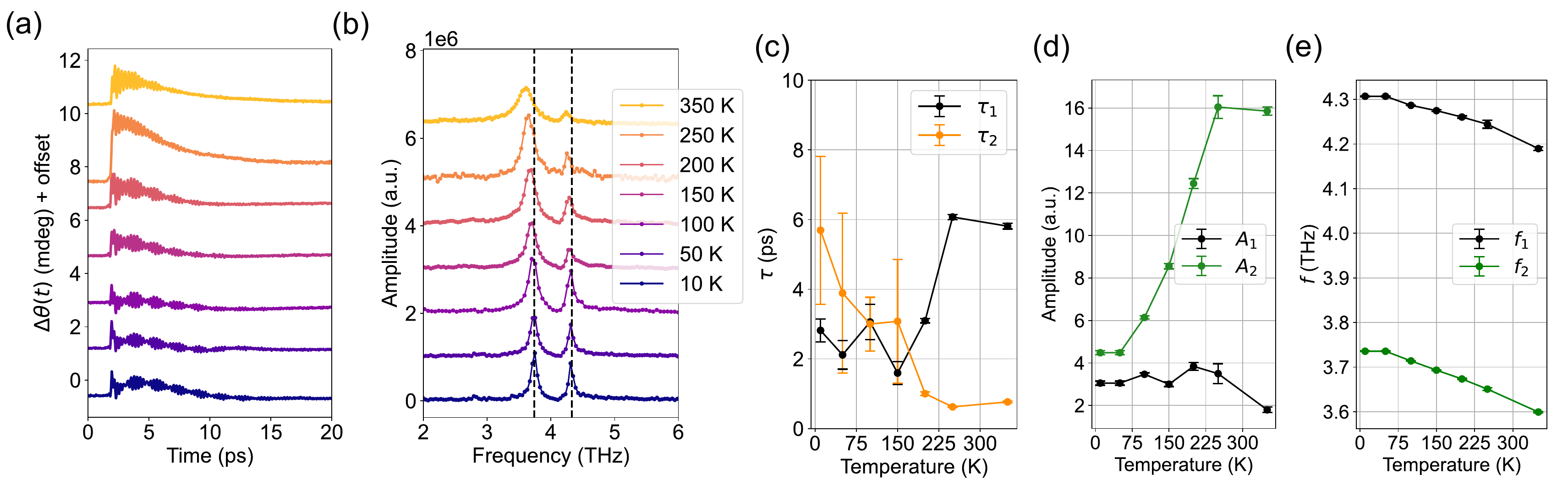}
\caption{\label{fig:Fig4} Temperature dependence of the 3.6 THz and 4.2 THz optical phonons. (a) $\Delta \theta(t)$ traces from 10 K to 350 K and (b) corresponding FFTs. At higher temperatures, the phonons broaden in linewidth, redshift, and have shortened lifetime, in agreement with previous Raman studies. (c) Fit time scales $\tau_1$ and $\tau_2$ after fitting the background to a two-temperature model described in Eqn. 4. (d) Amplitudes and (e) frequencies of the two phonons, obtained by fitting to two damped sinusoids after background subtraction. }
\end{figure*}

Lastly, we measure $\Delta \theta(t)$ as a function of temperature from 10 K up to 350 K in a cryostat. In this configuration, the maximum attainable $H_{app}$ was $\approx$ 500 G, which we found was not sufficient to resolve spin oscillations. Instead, we focus on the optical phonon oscillations, showing $\Delta \theta(t)$ traces in Fig. \ref{fig:Fig4}(a) and FFTs in \ref{fig:Fig4}(b). (Full time traces showing no spin oscillations are shown in the SI.) 

To analyze the temperature dependence of the phonons, we fit the background, then subtract it off and fit the residual oscillations to two damped sinusoids (see SI). We find that this two-step process is more robust than fitting the background and oscillations simultaneously. At low temperatures, modeling the background by a sum of decaying exponentials does not yield a satisfactory fit. Therefore, we adapt a phenomenological model developed to fit pump-probe reflectivity in Ref. \onlinecite{DjordjevicPhysStatSolidi2006}:

\begin{equation}
\begin{split}
 \Delta T_{el}(t) &= \Delta T_{e,max} \mathrm{erf}\left( \frac{t}{\tau_{ee}}\right)e^{-t/\tau_{el}},\\  \Delta T_{l}(t) &= \Delta T_{l,max} \mathrm{erf}\left( t/\tau_0\right) \left( 1-e^{-\frac{t+\tau_0}{\tau_{el}}} \right), \\ \Delta R/R(t) &= A \Delta T_{e}(t) + B \Delta T_{l}(t),
\end{split}
\end{equation}

Eqn. 6 approximates a two-temperature model in which $\Delta T_{el}(t)$ is the electron temperature, $\Delta T_{l}(t)$ is the lattice temperature, $\tau_{ee}$ is the electron-electron equilibration time, $\tau_{el}$ is the electron-phonon relaxation time, $\tau_0 \sim$ 100 fs is the pump pulse width, and $A$ and $B$ are fitting amplitudes relating temperature changes to optical response. Since $\Delta \theta(t)$ is not identical to reflectivity, we fit to:

\begin{equation}
\begin{split}
\Delta \theta(t) &= \Delta \theta_{K,1}(t) + \Delta \theta_{K,2}(t) \\
&= C_1 \mathrm{erf} \left( \frac{t}{\tau_{1}}\right)e^{-t/\tau_{2}} + C_2 \mathrm{erf} \left(\tau/\tau_0\right) \left( 1-e^{-\frac{\tau+\tau_0}{\tau_2}}\right)
\end{split}
\end{equation}

without explicitly identifying $\tau_1$ and $\tau_2$ with $\tau_{ee}$ and $\tau_{el}$. Fitting results from the electron and lattice time scales are shown in Fig. \ref{fig:Fig4}(c). $\tau_{1}$ increases with increasing $T$, while $\tau_{2}$ decreases. Although the background is well-fit by this model (see SI for details), $\tau_{1}$ is too long to be straightforwardly associated with $\tau_{ee}$, which is typically $<$ 1 ps. However, $\tau_{2}$ is on the same order as typical $\tau_{el}$, and the increase of $\tau_2$ with $T$ may correspond to the known increase of $\tau_{el}$ with increasing electron temperature \cite{MansartPRB2013}. Unlike Ref. \onlinecite{BossiniPRB2021}, we do not observe a sublattice demagnetization time scale that changes dramatically above $T_N$, which suggests that the 3.6 and 4.2 THz phonons are unrelated to the spin oscillations. 

After fitting the background, we fit the residual optical phonon oscillations to two damped sinusoids as in Eqn. 2: $\Delta \theta(t) = A_1 e^{-\alpha_1 t}\sin(2\pi f_1 t + \phi_1) + A_2 e^{-\alpha_2 t}\sin(2\pi f_2 t + \phi_2)$. The phonon fit amplitudes $A_1, A_2$ and frequencies $f_1, f_2$ are shown in Fig. \ref{fig:Fig4}(d) and (e). The frequencies monotonically decrease with increasing $T$, in agreement with previous Raman studies \cite{ZhangJRamanSpec2020}. We find that the amplitude of the 4.2 THz phonon $A_1$ decreases slightly with increasing $T$, while $A_2$ (the 3.6 THz phonon) increases until it remains constant above 250 K. 

In conclusion, we used TR-MOKE to identify a magnon mode and two optical phonon modes in the antiferromagnetic semiconductor MnTe. Both the magnon and the optical phonon can be suppressed or enhanced by changing pump polarization, demonstrating a non-thermal origin. We continued to observe field-dependent oscillations above $T_N$, at least up to 335 K, which may reflect the persistence of magnetic anisotropy and coupling to short-range order above $T_N$ and will be the subject of future work. Our results provide new insight into this unique magnetic system, and will be important as MnTe continues to be studied for spintronic and altermagnetic applications.

\begin{acknowledgments}
The authors thank Prof. Jay Kikkawa for helpful discussions. I.G. was mainly sponsored by the Army Research Office under the grant No. W911NF-20-2-0166. The construction of the pump-probe setup was supported by the Air Force Office of Scientific Research under award no. FA9550-22-1-0410. Q.D. was supported  by the NSF EPM program under grant no. DMR-2213891 and the Vagelos Institute of Energy Science and Technology graduate fellowship. Q.T. was support by the US Office of Naval Research (ONR) through the grant N00014-24-1-2064. The work at ORNL was supported by the U. S. Department of Energy (DOE), Office of Science, Basic Energy Sciences (BES), Materials Sciences and Engineering Division (growth and materials characterization).
\end{acknowledgments}

\section*{Data Availability Statement}
The data that support the findings of this study are available from the corresponding author upon reasonable request.

\section*{Author Declarations}
\subsection*{Conflict of Interest}
The authors have no conflicts to disclose.

\section*{Author Contributions}
\textbf{Isaiah Gray:} Conceptualization (equal); Data curation (lead); Formal analysis (lead); Methodology (equal); Validation (lead); Software (supporting); Visualization (lead); Writing - original draft (lead); Writing - review and editing (equal). \textbf{Qinwen Deng}: Data curation (supporting); Formal analysis (supporting); Methodology (equal); Software (equal). \textbf{Qi Tian:} Methodology (equal); Software (equal). Michael Chilcote: Resources (equal). \textbf{J. Steven Dodge:} Writing: review and editing (supporting). \textbf{Matthew Brahlek:} Resources (equal); Writing - review and editing (equal). \textbf{Liang Wu:} Conceptualization (equal); Funding Acquisition (lead); Methodology (equal); Project Administration (lead), Supervision (lead), Writing - review and editing (equal).

\bibliography{aipsamp}% Produces the bibliography via BibTeX.

\bibliography{aipsamp}% Produces the bibliography via BibTeX.

\pagebreak

%\renewcommand{\thefigure}{S\arabic{figure}}

%\documentclass[aps,prl,twocolumn,groupedaddress]{revtex4}

% You should use BibTeX and apsrev.bst for references
% Choosing a journal automatically selects the correct APS
% BibTeX style file (bst file), so only uncomment the line
% below if necessary.
%\usepackage{graphicx}
%\usepackage{epstopdf}
%\usepackage{color}
%\begin{document}
%\preprint{}
\begin{center}
\textbf{\large Supplemental Materials for "Time-resolved magneto-optical Kerr effect in the altermagnet candidate MnTe"}
\end{center}

\setcounter{equation}{0}
\setcounter{figure}{0}
\setcounter{table}{0}
\setcounter{page}{1}

\makeatletter
\renewcommand{\theequation}{S\arabic{equation}}
\renewcommand{\thefigure}{S\arabic{figure}}
\renewcommand{\thepage}{S\arabic{page}}
\renewcommand{\bibnumfmt}[1]{[S#1]}
\renewcommand{\citenumfont}[1]{S#1}

\section{Magnetic field dependence of optical phonons}

\begin{figure*}[h]
\includegraphics[width=\textwidth]{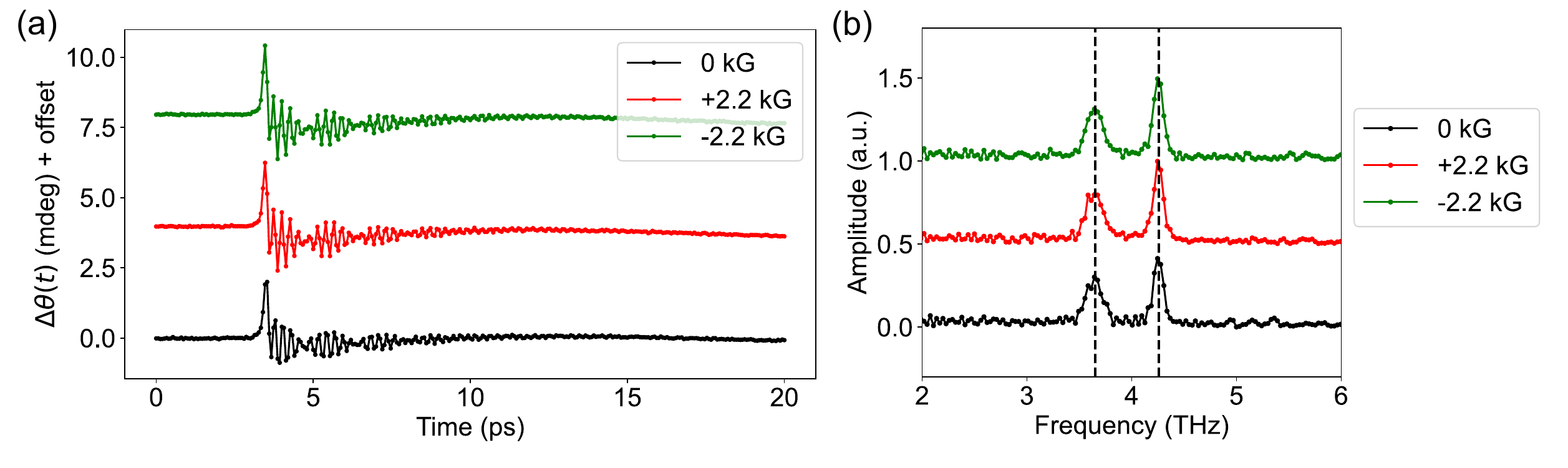}
\caption{\label{fig:S1} (a) $\Delta \theta(t)$ time traces and (b) FFTs showing optical phonons at 295 K at $H_{app}$ = +2.2 kG, 0 kG, and -2.2 kG out-of-plane.}
\end{figure*}

In Fig. \ref{fig:S1}, we show short scans of $\Delta \theta(t)$ scans at $H_{app}$ = +2.2 kG, 0 kG, and -2.2 kG out-of-plane, with corresponding FFTs. The oscillations repeat nearly point-for-point, showing no change with field. Therefore, we ascribe this signal purely to optical phonons and not phonon-induced magnons. Vertical dashed lines are drawn at $f$ = 3.65 THz and $f$ = 4.26 THz in \ref{fig:S1}(b) as a visual aid to show that the peaks do not shift. 

\section{Long scans at low temperature}

\begin{figure*}
\includegraphics[width=0.9\textwidth]{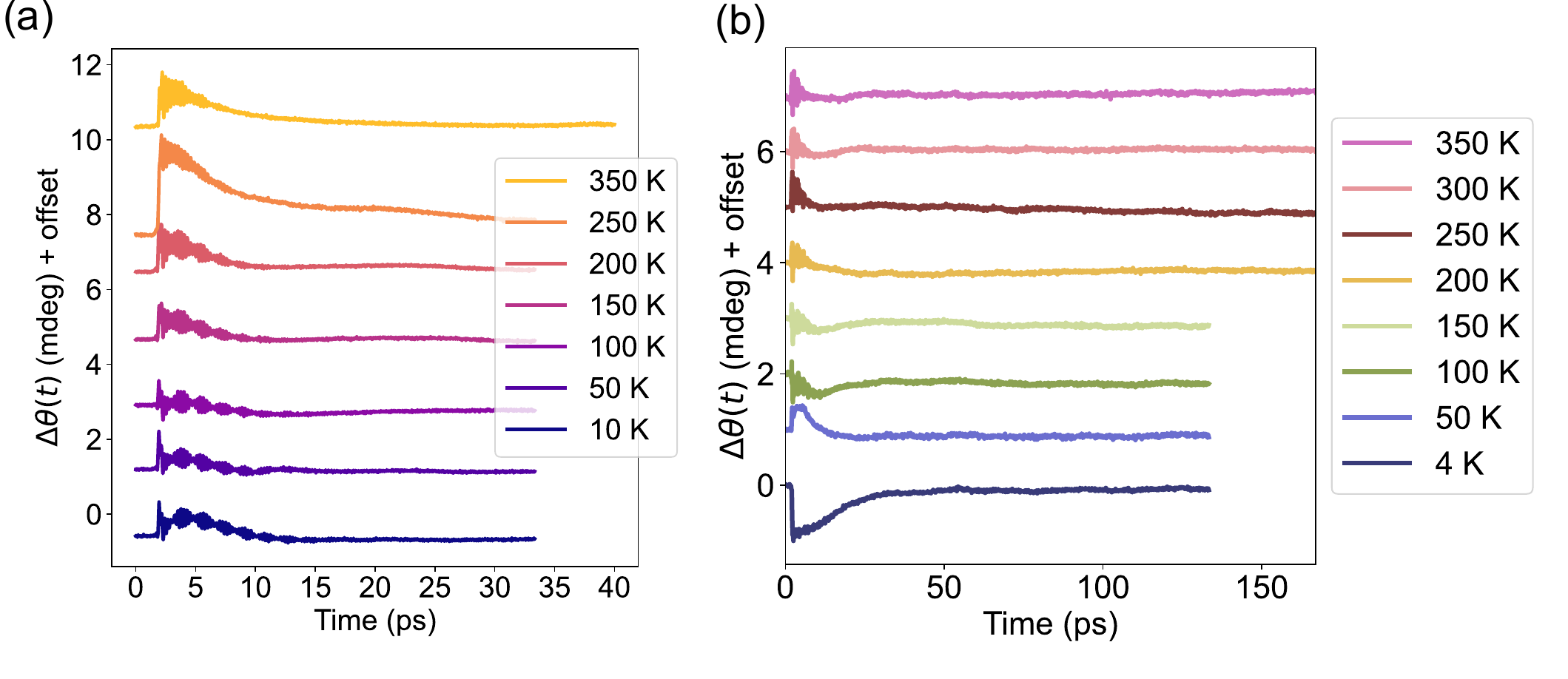}
\caption{\label{fig:S2} (a) Full data set from Fig. 4(a) of the main text, with 35 ps time extent. (b) A separate temperature-dependent run with 133 ps time extent, showing no magnon oscillations.}
\end{figure*}

Here, we show long scans at low temperature at $H_{app}$ = 500 G, demonstrating no magnon oscillations. In Fig. \ref{fig:S2}(a), we show the full data set from Fig. 4 of the main text, which restricted the time extent to 20 ps so that the optical phonon oscillations could be more easily seen. In Fig. \ref{fig:S2}(b), we present a separate set of $\Delta \theta(t)$ scans with 133 ps time extent on the same sample, also with $H_{app}$ = 500 G. Within experimental sensitivity ($\sim 1 \mu$V), $\Delta \theta(t)$ is constant after electron-lattice relaxation.

\section{Fitting TR-MOKE background at low temperatures}

\begin{figure*}
\includegraphics[width=\textwidth]{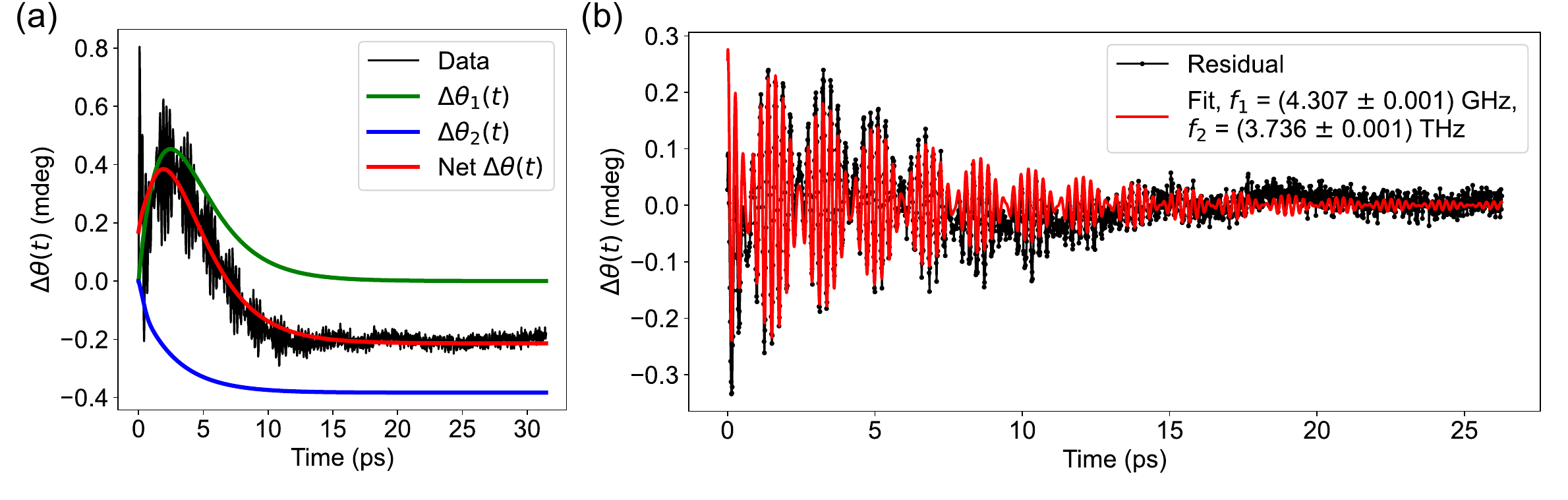}
\caption{\label{fig:S3}Fitting 10 K data from Fig. 4 of the main text. (a) Fitting the background to Eqn. 6. (b) Fitting the residual oscillations (with background subtracted) to two damped sinusoids.}
\end{figure*}

In this section, we detail the fitting process for the low-temperature data in Fig. 4 of the main text. We show the fitting at 10 K; other temperatures are processed in exactly the same way. Fig. \ref{fig:S3}(a) shows the raw data at 10 K, the individual contributions from the best fit $\Delta \theta_{1}(t)$ and $\Delta \theta_{2}(t)$ from Eqn. 6, and their sum $\Delta \theta_{net}(t)$. In Fig. \ref{fig:S3}(b), we subtract the best-fit background from the data and fit the residuals to $\Delta \theta(t) = A_1 e^{-\alpha_1 t}\sin \left(2 \pi f_1 t + \phi_1 \right) + A_2 e^{-\alpha_2 t}\sin \left( 2 \pi f_2 t + \phi_2 \right)$.

\section{Details of fitting magnetic field and pump polarization dependence}

\begin{figure*}
\includegraphics[width=0.9\textwidth]{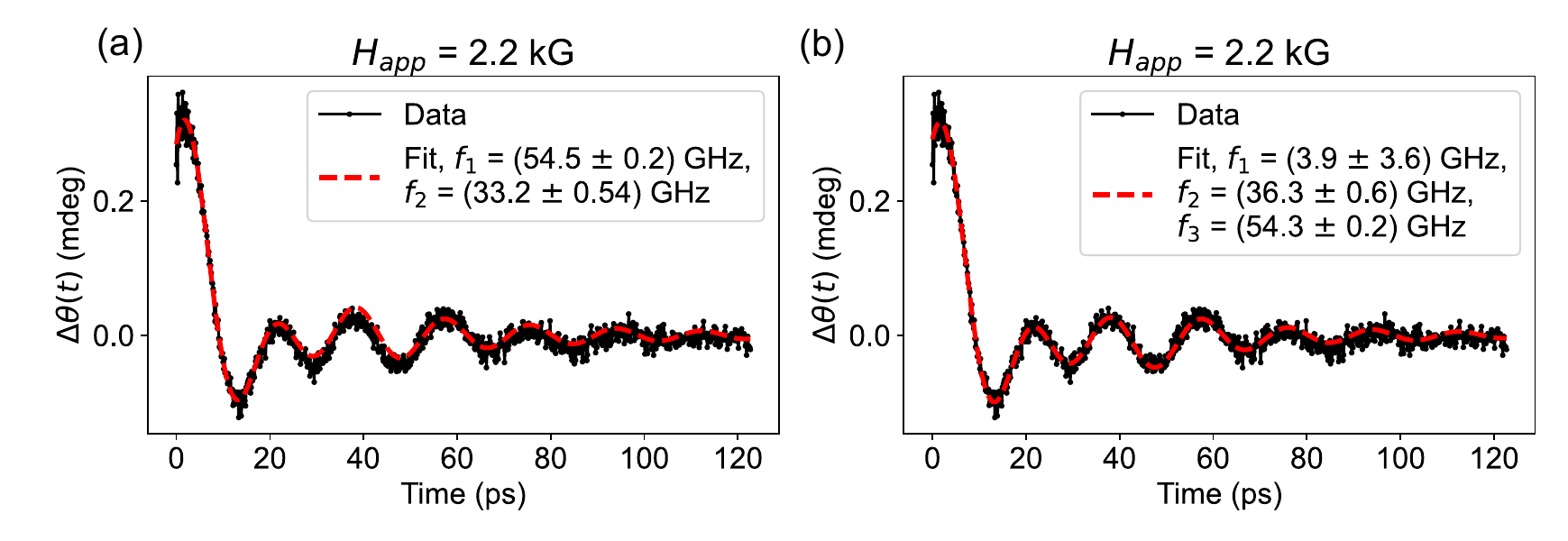}
\caption{\label{fig:S4}Fitting the data at $H_{app}$ = +2.2 kG from Fig. 2(a) of the main text. At this pump angle (225$^\circ$), three damped sines do not provide an improved fit compared to two damped sines.}
\end{figure*}

Here we provide details of the fitting of the spin oscillations as well as the optical phonon oscillations in Figs. 1 and 2 of the main text. In Fig. \ref{fig:S4}, we show an example of fitting the magnetic field dependence at fixed 225$^\circ$ pump angle in Fig. 2(a) of the main text. We fit to two damped sinusoids in Fig. \ref{fig:S4}(a) and three damped sinusoids in \ref{fig:S4}(b). In this set, a third oscillation is not well-resolved ($f_1 = 3.9 \pm 3.6$ GHz) and the quality of the fit is not improved by including it. Data at other values of $H_{app}$ in Fig. 2(a) shows similar behavior.

\begin{figure*}
\includegraphics[width=0.75\textwidth]{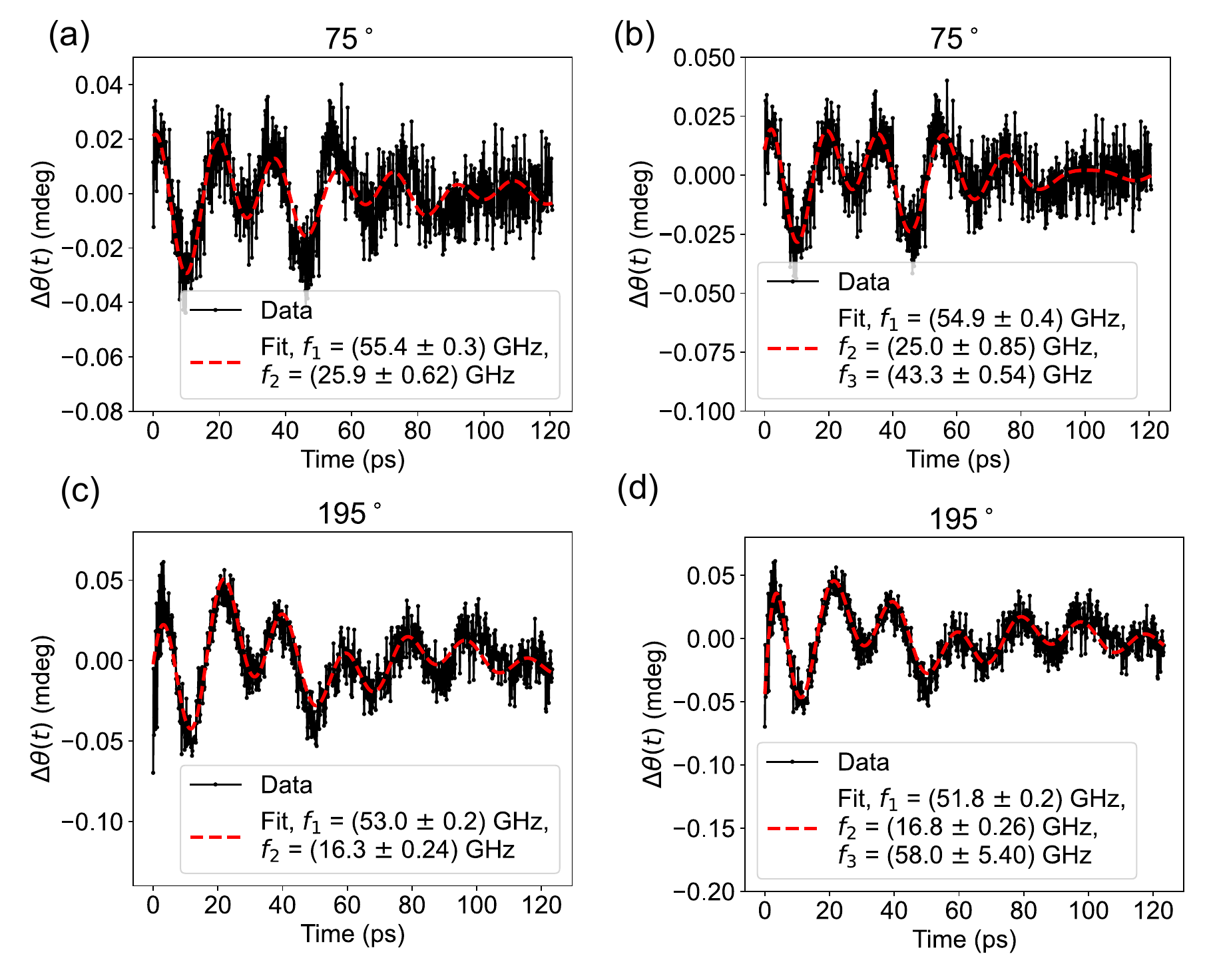}
\caption{\label{fig:S5}Fitting polarization dependence from Fig. 1 and 2 of the main text. (a,b) Fitting the spin oscillation data at 75$^\circ$ pump angle to two and three damped sinusoids, respectively. (c,d) Repeating the process at 195$^\circ$ pump angle. While three sinusoids fit better than two at 75$^\circ$, no noticable improvement is seen at 195$^\circ$. }
\end{figure*}

We then fit the pump polarization dependence in Fig. \ref{fig:S5}. We fit the spin oscillation at 75$^\circ$ pump angle to two and three sinusoids in \ref{fig:S5}(a) and (b), respectively, and compare to similar fits of the oscillations at 195$^\circ$ pump angle in (c) and (d). In the 75$^\circ$ data, three frequencies, fitting to 25, 43, and 55 GHz, are better than two, while in the 195$^\circ$ data, the fits are nearly identical and $f_3 = 58 \pm 5$ GHz is not well-distinguished from $f_1 = 51.8 \pm 0.2$ GHz. This suggests that there are at least three frequencies - likely one magnon and two acoustic phonons - with different pump angle dependence. 

%Finally, we fit the optical phonon pump dependence from Fig. 1(c) of the main text to two damped sinusoids, showing an example at 195$^{\circ}$ in Fig. \ref{fig:S5}(e). We plot the resulting fit amplitudes $A_1$ and $A_2$ as a function of pump angle in \ref{fig:S5}(f), and in turn fit those amplitudes to $\sin (3 \theta)$ to test for three-fold symmetry. While $A_1$ is inconclusive, $A_2$ indeed exhibits three-fold symmetry.

\section{Fitting $\Delta \theta (t)$ above $T_N$}

\begin{table}
\caption{\label{tab:table3} Fit parameters for the field-dependent data in Fig. 3(b) and (c) of the main text. }
\begin{ruledtabular}
\begin{tabular}{ccccccc}
 & $f_1$ (GHz) &  $f_2$ (GHz) & $f_3$ (GHz) & $A_1$ (mdeg) & $A_2$ (mdeg) & $A_3$ (mdeg)  \\
 \hline
295 K, +2.2 kG & 13.7(5) & 33.8(4) & 59.6(1.5) & 0.19(1) & 0.36(5) & 0.25(6) \\
295 K, -2.2 kG & N/A & 33.8(2) & 64(1) & N/A & 0.69(5) & 1.0(4) \\ 

325 K, +2.2 kG & N/A & 31.4(4) & 58(2) & N/A & 0.20(3) & 1.4(5) \\

325 K, 0 kG & 15.3(3) & 34.9(3) & N/A & 0.07(1) & 0.33(4) & N/A \\

325 K, -2.2 kG & 18(2) & 38.3(4) & 59.1(4) & 0.6(1) & 0.56(5) & 0.68(7) \\

\end{tabular}
\end{ruledtabular}

\end{table} 

In this section, we show the fit parameters for the field-dependent data in Fig. 3(b) and (c) of the main text. We follow the same fitting procedure as in the rest of the text: we fit the raw data to a background of a decaying exponential plus a linear function, subtract off the background, then fit the remaining oscillations to three exponentially damped sinusoids. Resulting frequencies and amplitudes for each sinusoid are shown in Table 1, with uncertainties on the last significant digit given in parentheses. As in Fig. 2 of the main text, the data at 295 K and 0 kG did not exhibit sufficient oscillation amplitude to allow a consistent fit.

We find that at both 295 K and 325 K, the $f_3$ mode (between 58 and 64 GHz) appears only when $H_{app} \neq$ 0, whereas the $\sim$15 GHz and 35 GHz modes are also present at 325 K and 0 kG. Note that no constraints on the frequency values were imposed in the fitting, so the presence or absence of the 55 GHz mode is not an artifact of bounding the frequencies.

\end{document}